\documentclass[preprint,prd,nofootinbib,tightenlines,groupedaddress,amsmath,amssymb]{revtex4}

\usepackage{bm}
\usepackage{color}
\usepackage{graphicx}
\usepackage[hypertex]{hyperref}
\usepackage{multirow}

\begin{document}
\baselineskip=15pt \parskip=3pt

\vspace*{3em}

\title{Exploring X-Ray Lines as Scotogenic Signals}

\author{Gaber Faisel,$^{1,2}$  Shu-Yu Ho,$^1$  and Jusak Tandean$^1$ }
\affiliation{${}^{1}$Department of Physics and Center for Theoretical  Sciences,
National Taiwan University, \\ Taipei 106, Taiwan \vspace{3pt} \\
${}^{2}$Egyptian Center for Theoretical Physics,
Modern University for Information and Technology,\\ Cairo, Egypt \\
$\vphantom{\bigg|_{\bigg|}^|}$}



\begin{abstract}
We consider some implications of X-ray lines from certain astronomical
objects as potential effects of dark matter decay in the context of the scotogenic model,
where neutrinos acquire mass radiatively via one-loop interactions with dark matter.
As an example, we focus on the 3.5 keV line recently detected in the X-ray spectra of
galaxy clusters, assuming that it stands future scrutiny.
We explore the scenario in which the line originates from the slow decay of fermionic
dark matter in the model.
After obtaining a~number of benchmark points representing the parameter space consistent
with the new data and various other constraints, we make predictions on several
observables in leptonic processes.
They include the effective Majorana mass in neutrinoless double-beta decay,
the sum of neutrino masses, and the rate of flavor-changing decay $\mu\to e\gamma$,
as well as the cross sections of $e^+e^-$ collisions into final states containing
nonstandard particles in the model.
These are testable in ongoing or future experiments and thus offer means to probe
the scotogenic scenario studied.
\end{abstract}

\maketitle

\section{Introduction\label{intro}}

Earlier this year two different collaborations~\cite{Bulbul:2014sua,Boyarsky:2014jta}
independently reported the detection of a~weak unidentified emission line at an energy
of \,$\sim3.5$\,keV\, in the X-ray spectra of a number of galaxy clusters and
the Andromeda galaxy.
More recently, the second group arrived at an almost identical result after examining
the dataset of the Milky Way center~\cite{Boyarsky:2014ska}.
On the other hand, there have been analyses~\cite{Riemer-Sorensen:2014yda} that
questioned these findings.
The second of the groups that had made the positive announcements has also responded
to some of the criticisms of their results~\cite{Boyarsky:2014paa}.

Pending a consensus on whether the signal exists or not, one can therefore adopt
the position that it does.
In that case, it may be an intriguing hint of new physics beyond the standard model (SM)
of particle physics, being compatible with the characteristics of a line originating from
the decay of a nonstandard particle~\cite{Bulbul:2014sua,
Boyarsky:2014jta,Boyarsky:2014ska,f2fg,Frandsen:2014lfa,Modak:2014vva}.\footnote{The 3.5 keV
X-ray may alternatively be produced by dark-matter annihilation~\cite{Frandsen:2014lfa,ann}.}

In this paper we work along a~similar line within the context of the scotogenic model,
which was invented by Ma~\cite{Ma:2006km}.
Among the salient features of the model is the intimate link between its neutrino and dark
matter (DM) sectors, as the neutrinos acquire mass radiatively via one-loop interactions with
new particles, the lightest of which can act as weakly-interacting massive particle~DM.
The nonstandard ingredients of the model consist merely of a~scalar doublet,
$\eta$, and three singlet Majorana fermions, $N_{1,2,3}$, all of which are odd under an~exactly
conserved $Z_2$ symmetry.   In contrast, all of the SM particles are $Z_2$ even.
This symmetry prevents neutrino mass from being generated at tree level and ensures
the DM stability.
Here we assume that the DM is composed of nearly degenerate $N_1$ and~$N_2$, with the latter
being the more massive, and the slow decay \,$N_2\to N_1\gamma$\, is responsible for
the observed X-ray line.
Various other aspects of the model have been explored in the literature~\cite{Kubo:2006yx,
Sierra:2008wj,Schmidt:2012yg,Ho:2013hia,Ho:2013spa}.

Previously, there was already a study on the 3.5 keV line within the same
model~\cite{Modak:2014vva}.
However, the choices made therein for the Yukawa couplings of the new particles correspond
to a zero value of the neutrino-mixing parameter $\sin\theta_{13}$ which has now been
established to be nonnegligible~\cite{pdg}.
More importantly, various constraints from the observed DM relic abundance, neutrino
oscillation data, and experimental limits on flavor-violating decays such as
\,$\mu\to e\gamma$\, were not addressed in~Ref.\,\cite{Modak:2014vva}, but they
significantly restrict the viable parameter space of the model.
In our analysis below we take into account these and other relevant factors carefully.
Moreover, from the allowed parameter space we make predictions on several
observables.
One of them is the effective Majorana mass that can be probed in ongoing and planned
searches for neutrinoless double-beta decay.
Another interesting quantity is the sum of neutrino masses that can be compared with
numbers inferred from upcoming cosmological measurements with increased precision.
Also pertinent are the rates of the loop-induced Higgs boson decays \,$h\to\gamma\gamma$\,
and \,$h\to\gamma Z$,\, which receive scotogenic contributions and
are already under investigation at the LHC.
In addition, we predict the cross sections of electron-positron scattering into final states
involving scotogenic particles that can be tested at next-generation $e^+e^-$ colliders.

The organization of the paper is as follows.
The next section gives a description of the relevant Lagrangians for the new particles
and the parameters associated with the neutrino masses.
In~Section~\ref{constraints}, we look at a number of constraints on the couplings and
masses of the new particles in the particular scenario of interest.
Upon scanning the parameter space of the model, we present some sets of benchmark points
representing parts of the viable regions.
In~Section~\ref{predictions}, we provide predictions on several quantities that can be
tested in ongoing and future experiments.
We give our conclusions in~Section~\ref{summary}.

\section{Lagrangians, couplings, and masses\label{interactions}}

The Lagrangian describing the interactions of the scalar particles in the scotogenic
model with one another and the gauge bosons is
\begin{eqnarray} \label{L}
{\cal L} \,\,=\,\, ({\cal D}^\varsigma\Phi)^\dagger\,{\cal D}_\varsigma\Phi \,+\,
({\cal D}^\varsigma\eta)^\dagger\,{\cal D}_\varsigma\eta \;-\; {\cal V} ~,
\end{eqnarray}
where ${\cal D}_\varsigma$ denotes the covariant derivative containing the SM gauge fields,
the potential~\cite{Ma:2006km}
\begin{eqnarray} \label{potential}
{\cal V} &\,=&\, \mu_1^2\,\Phi^\dagger \Phi \,+\, \mu_2^2\,\eta^\dagger\eta \,+\,
\mbox{$\frac{1}{2}$}\lambda_1^{}(\Phi^\dagger \Phi)^2 \,+\,
\mbox{$\frac{1}{2}$}\lambda_2^{}(\eta^\dagger\eta)^2
\nonumber \\ && +\;
\lambda_3^{}(\Phi^\dagger \Phi)(\eta^\dagger\eta) \,+\,
\lambda_4^{}(\Phi^\dagger\eta)(\eta^\dagger \Phi) \,+\,
\mbox{$\frac{1}{2}$}\lambda_5^{}\bigl[ (\Phi^\dagger\eta)^2+(\eta^\dagger\Phi)^2\bigr] ~,
\end{eqnarray}
and after electroweak symmetry breaking
\begin{eqnarray}
\Phi \,\,=\, \left(\!\begin{array}{c} 0 \vspace{1pt} \\
\frac{1}{\sqrt2}_{\vphantom{o}}(h+v) \end{array}\! \right) , \hspace{7ex}
\eta \,\,=\, \left(\!\begin{array}{c} H^+ \vspace{1pt} \\
\frac{1}{\sqrt2}_{\vphantom{o}}({\cal S}+i{\cal P}) \end{array}\! \right) ,
\end{eqnarray}
with $h$ being the physical Higgs boson and $v$ the vacuum expectation value (VEV) of $\Phi$.
Because of the $Z_2$ symmetry, the VEV of $\eta$ is zero.
The masses of $\cal S$, $\cal P$, and $H^\pm$ are then given by
\,$m_{\cal S}^2=m_{\cal P}^2+\lambda_{5\,}^{}v^2=
\mu_2^2+\frac{1}{2}(\lambda_3^{}+\lambda_4^{}+\lambda_5^{})v^2$\,
and \,$m_H^2=\mu_2^2+\frac{1}{2}\lambda_3^{}\,v^2$.\,
We make the usual assumption that $\lambda_5$ is small~\cite{Kubo:2006yx},
\,$|\lambda_5|\ll|\lambda_3+\lambda_4|$,\, implying that $m_{\cal S,P}^{}$ are nearly degenerate,
\,$|m_{\cal S}^2-m_{\cal P}^2|=|\lambda_5|v^2\ll m_{\cal S}^2\simeq m_{\cal P}^2$.\,

The Lagrangian for the masses and interactions of the new singlet fermions $N_k$ is
\begin{eqnarray} \label{LN}
{\cal L}_N^{} \,\,=\,\,
-\mbox{$\frac{1}{2}$} M_k^{}\,\overline{N_k^{\rm c}}\,P_R^{} N_k^{} \,+\,
{\cal Y}_{rk}^{} \Bigl[ \bar\ell_r^{} H^- \,-\,
\mbox{$\frac{1}{\sqrt2}$}\,\bar\nu_r^{}\,({\cal S}-i {\cal P}) \Bigr] P_R^{} N_k^{}
\;+\; {\rm H.c.} ~,
\end{eqnarray}
where $M_k$ represent their masses, \,$k,r=1,2,3$\, are summed over,
the superscript c refers to charge conjugation, \,$P_R=\frac{1}{2}(1+\gamma_5)$,\,
and \,$\ell_{1,2,3}=e,\mu,\tau$.\,
The Yukawa couplings of $N_k$ form the matrix
\begin{eqnarray} \label{yukawa}
{\cal Y} \,\,=\, \left(\begin{array}{ccc} Y_{e1} & Y_{e2} & Y_{e3} \vspace{2pt} \\
Y_{\mu1} & Y_{\mu2} & Y_{\mu3} \vspace{2pt} \\ Y_{\tau1} & Y_{\tau2} & Y_{\tau3}
\end{array}\right) ,
\end{eqnarray}
where \,$Y_{\ell_r k}={\cal Y}_{r k}^{}$.\,

The neutrinos get mass radiatively through one-loop diagrams with internal
$\cal S$, $\cal P$, and~$N_k$.
The mass eigenvalues $m_j^{}$ are given by~\cite{Ma:2006km}
\begin{eqnarray} \label{UMU} & \displaystyle
{\rm diag}\bigl(m_1^{},m_2^{},m_3^{}\bigr) \,\,=\,\, {\cal U}^\dagger{\cal M}_\nu\,{\cal U}^* \,,
& \\ \label{Mnu} & \displaystyle
{\cal M}_\nu^{} \,\,=\,\, {\cal Y}\, {\rm diag}(\Lambda_1,\Lambda_2,\Lambda_3)\,
{\cal Y}^{{\rm T}^{\vphantom{\displaystyle|}}}  \,,
& \\ \label{lambdak} & \displaystyle
\Lambda_k^{} \,\,=\,\, \frac{\lambda_{5\,}^{}v^2}{16\pi^2M_k^{}}\Biggl[ \frac{M_k^2}{m_0^2-M_k^2}
+ \frac{2 M_k^4\,\ln\bigl(M_k^{}/m_0^{}\bigr)}{\bigl(m_0^2-M_k^2\bigr)\raisebox{1.7pt}{$^{\!2}$}}
\Biggr]^{\vphantom{\int^|}} \,, \hspace{7ex}
m_0^2 \,\,=\,\, \mbox{$\frac{1}{2}$}\bigl(m_{\cal S}^2 + m_{\cal P}^2\bigr) ~, &
\end{eqnarray}
where $\,\cal U$ is the Pontecorvo-Maki-Nakagawa-Sakata (PMNS~\cite{pmns}) unitary matrix and
the formula for $\Lambda_k^{}$ applies to
the \,$m_0^{}\simeq m_{\cal S}^{}\simeq m_{\cal P}^{}$\, case.
For $\cal U$, we choose the PDG parametrization~\cite{pdg}
\begin{eqnarray} \label{U}
{\cal U} &\,=\,& \tilde u\;{\rm diag}\bigl(e^{i\alpha_1/2},e^{i\alpha_2/2},1\bigr) \,, \\
\tilde u &\,=& \left(\begin{array}{ccc}
 c_{12\,}^{}c_{13}^{} & s_{12\,}^{}c_{13}^{} & s_{13}^{}\,e^{-i\delta^{\vphantom{|}}}
\vspace{3pt} \\
-s_{12\,}^{}c_{23}^{}-c_{12\,}^{}s_{23\,}^{}s_{13}^{}\,e^{i\delta} & ~~
 c_{12\,}^{}c_{23}^{}-s_{12\,}^{}s_{23\,}^{}s_{13}^{}\,e^{i\delta} ~~ & s_{23\,}^{}c_{13}^{}
\vspace{3pt} \\
 s_{12\,}^{}s_{23}^{}-c_{12\,}^{}c_{23\,}^{}s_{13}^{}\,e^{i\delta} &
-c_{12\,}^{}s_{23}^{}-s_{12\,}^{}c_{23\,}^{}s_{13}^{}\,e^{i\delta} & c_{23\,}^{}c_{13}^{}
\end{array}\right) ,
\end{eqnarray}
where \,$\delta\in[0,2\pi]$\, and \,$\alpha_{1,2}^{}\in[0,2\pi]$\, are the Dirac and Majorana
$CP$-violation phases, respectively, \,$c_{mn}^{}=\cos\theta_{mn}^{}\ge0$,\, and
\,$s_{mn}^{}=\sin\theta_{mn}^{}\ge0$.\,

The Yukawa couplings $Y_{\ell_r k}$ need to satisfy the relations in Eqs. (\ref{UMU})
and (\ref{Mnu}).
We adopt the solutions employed in Ref.\,\cite{Ho:2013spa}, namely,
\begin{eqnarray} \label{123} & \displaystyle
Y_{e1}^{} \,\,=\,\, \frac{-c_{12\,}^{}c_{13}^{}\,Y_1^{}}
{c_{12\,}^{}c_{23\,}^{}s_{13\,}^{}e^{i\delta}-s_{12\,}^{}s_{23}^{}} ~, ~~~~~~~ &
Y_{\mu1}^{} \,\,=\,\, \frac{c_{12\,}^{}s_{23\,}^{}s_{13\,}^{}e^{i\delta}+s_{12\,}^{}c_{23}^{}}
{c_{12\,}^{}c_{23\,}^{}s_{13\,}^{}e^{i\delta}-s_{12\,}^{}s_{23}^{}}\; Y_1^{} ~,
\nonumber \\ & \displaystyle
Y_{e2}^{} \,\,=\,\, \frac{-s_{12\,}^{}c_{13}^{}\,Y_2^{}}
{s_{12\,}^{}c_{23\,}^{}s_{13\,}^{}e^{i\delta}+c_{12\,}^{}s_{23}^{}} ~, ~~~~~~~ &
Y_{\mu2}^{} \,\,=\,\, \frac{s_{12\,}^{}s_{23\,}^{}s_{13\,}^{}e^{i\delta}-c_{12\,}^{}c_{23}^{}}
{s_{12\,}^{}c_{23\,}^{}s_{13\,}^{}e^{i\delta}+c_{12\,}^{}s_{23}^{}}\; Y_2^{} ~,
\nonumber \\ & \displaystyle \vphantom{|^{\int_\int^\int}}
Y_{e3}^{} \,\,=\,\, \frac{s_{13}^{}\,Y_3^{}}{c_{23\,}^{}c_{13\,}^{}e^{i\delta}} ~, \hspace{17ex} &
Y_{\mu3}^{} \,\,=\,\, \frac{s_{23\,}^{}Y_3^{}}{c_{23}^{}} ~,
\end{eqnarray}
corresponding to the neutrino mass eigenvalues
\begin{eqnarray} \label{m1m2m3}
m_1^{} \,=\, \frac{\Lambda_{1\,}^{}Y_{e1}^2\,e^{-i\alpha_1}}{c_{12\,}^2c_{13}^2} ~, \hspace{7ex}
m_2^{} \,=\, \frac{\Lambda_{2\,}^{}Y_{e2}^2\,e^{-i\alpha_2}}{s_{12\,}^2c_{13}^2} ~, \hspace{7ex}
m_3^{} \,=\, \frac{\Lambda_{3\,}^{}Y_3^2}{c_{13\,}^2 c_{23}^2} ~.
\end{eqnarray}
The requirement that $m_{1,2,3}^{}$ be real and nonnegative then implies
\begin{eqnarray} \label{alpha12} & \displaystyle
\alpha_1^{} \,\,=\,\, \arg\bigl(\Lambda_{1\,}^{}Y_{e1}^2\bigr) ~, \hspace{7ex}
\alpha_2^{} \,\,=\,\, \arg\bigl(\Lambda_{2\,}^{}Y_{e2}^2\bigr) ~, \hspace{7ex}
\arg\bigl(\Lambda_{3\,}^{}Y_3^2\bigr) \,\,=\,\, 0 ~.
\end{eqnarray}
These choices are consistent with the neutrino oscillation data,
including \,$\sin\theta_{13}\neq0$.\,

Information on the values of some of the neutrino parameters above is available from various
measurements.
A recent fit to the global data on neutrino oscillations~\cite{Capozzi:2013csa} yield
\begin{eqnarray}
\sin^2\theta_{12}^{} &\,=\,& 0.308\pm0.017 ~, ~~~~~~~
\sin^2\theta_{23}^{} \,=\, 0.437_{-0.023}^{+0.033} ~,
\nonumber \\
\sin^2\theta_{13}^{} &\,=\,& 0.0234_{-0.0019}^{+0.0020} ~, \hspace{10ex}
\delta/\pi \,=\, 1.39_{-0.27}^{+0.38} ~,
\nonumber \\
\delta m^2 &\,=\,& m_2^2-m_1^2 \,=\,
\left(7.54_{-0.22}^{+0.26}\right)\times10^{-5}\;{\rm eV}^2 ~,
\nonumber \\
\Delta m^2 &\,=\,& m_3^2-\mbox{$\frac{1}{2}$}\bigl(m_1^2+m_2^2\bigr) \,=\,
\bigl(2.43_{-0.06}^{+0.06}\bigr)\times10^{-3}\;{\rm eV}^2 ~. \label{nudata}
\end{eqnarray}
These belong to the normal hierarchy of neutrino masses
$\bigl(m_1^{}<m_2^{}<m_3^{}\bigr)$, which is preferred by the solutions in Eq.\,(\ref{123}).
In contrast to the well-determined squared-mass differences in Eq.\,(\ref{nudata}),
the absolute scale of the masses is still poorly known.
The latest tritium $\beta$-decay experiments have led to a cap on
the (electron based) antineutrino mass of \,$m_{\bar\nu_e}<2\;$eV~\cite{pdg}.
Indirectly, stronger bounds on the total mass \,$\Sigma_k^{}m_k^{}=m_1^{}+m_2^{}+m_3^{}$\,
can be inferred from cosmological observations.
Specifically, the Planck Collaboration extracted \,$\Sigma_k^{}m_k^{}<0.23$\,eV\, at 95\%\,CL
from cosmic microwave background (CMB) and baryon acoustic oscillation (BAO)
measurements~\cite{planck}.  Including additional observations can
improve this limit to \,$\Sigma_k^{}m_k^{}<0.18$\,eV~\cite{Riemer-Sorensen:2013jsa}.
On the other hand, there are also recent analyses that have turned up tentative indications
of bigger masses.
The South Pole Telescope Collaboration reported \,$\Sigma_k^{}m_k^{}=(0.32\pm0.11)$\,eV\,
from the combined CMB, BAO, Hubble constant, and Sunyaev-Zeldovich selected galaxy cluster
abundances dataset~\cite{Hou:2012xq}, compatible with the later finding
\,$\Sigma_k^{}m_k^{}=(0.36\pm0.10)$\,eV\, favored by
the Baryon Oscillation Spectroscopic Survey CMASS Data Release 11~\cite{Beutler:2014yhv}.
As for the Majorana phases $\alpha_1^{}$ and $\alpha_2^{}$, there is still no empirical
information available on their values.

\section{Constraints\label{constraints}}

One interpretation of the 3.5 keV X-ray line suggested in
Refs.~\cite{Bulbul:2014sua,Boyarsky:2014jta} is that it is the signature of a sterile neutrino
$\nu_s^{}$ of mass \,$m_{\nu_s^{}}\sim7$\,keV\, which serves as DM and decays into the photon
plus an active neutrino at a rate within the range
\,$4.8\times10^{-48}\lesssim\Gamma_{\nu_s^{}\to\nu\gamma}/m_{\nu_s^{}}\lesssim4.6\times10^{-47}$.\,
In our scotogenic scenario of interest, $N_1$ plays the role of cold DM and is only slightly less
massive than $N_2$ such that the decay \,$N_2\to N_1\gamma$\, proceeds very slowly and is
responsible for the line.
Furthermore, $N_2$ has a lifetime $\tau_{N_2}^{}$ that is longer than the age of
the Universe, $\tau_U^{}$, and hence contributes to the DM density $\rho_{\rm DM}^{}$ with
present-day fractional abundance $f_{N_2}^{}$.\,
The near degeneracy of $N_1$ and $N_2$ implies that
\,$f_{N_2}^{}\simeq\frac{1}{2}\,e^{-\tau_U^{}/\tau_{N_2}^{}}$,\, where the exponential factor
accounts for the depletion of $N_2$ after freeze-out time.
Since the flux $\Phi_\gamma^{}$ of the X-rays is proportional to the rate-to-mass ratios in
the two cases according to
\,$\Phi_\gamma^{}\propto\rho_{\rm DM}^{}\Gamma_{\nu_s^{}\to\nu\gamma}/m_{\nu_s^{}}=
\rho_{\rm DM}^{}f_{N_2}^{}\Gamma_{N_2^{}\to N_1^{}\gamma}/M_2^{}$\, \cite{Frandsen:2014lfa},
we can then require
\begin{eqnarray} \label{xrayconstraint}
9.6\times10^{-48} \,\,<\,\, \frac{\Gamma_{N_2^{}\to N_1^{}\gamma}}{M_1^{}}\,
e^{-\tau_U^{}/\tau_{N_2}^{}} \,\,<\,\, 9.2\times10^{-47} ~,
\end{eqnarray}
where \,$\tau_U^{}=4.36\times10^{17}$\,s\,~\cite{pdg}.

This radiative decay arises from loop diagrams with internal $\ell_k^\pm$ and $H^\mp$
and the photon attached to either one of the charged particles.
We derive its rate to be
\begin{eqnarray} \label{N2toN1g}
\Gamma_{N_2^{}\to N_1^{}\gamma}^{} \,\,=\,\, \frac{\alpha E_\gamma^3M_1^2}{64\pi^4 m_H^4}
\Biggl[\raisebox{2pt}{\footnotesize$\displaystyle\sum_k$}\,
{\rm Im}\bigl({\cal Y}_{k1\,}^{}{\cal Y}_{k2}^*\bigr)\,{\cal G}
\Biggl(\frac{M_1^2}{m_H^2},\frac{m_{\ell_k}^2}{m_H^2}\Biggr) \Biggr]^2 ,
\end{eqnarray}
where \,$E_\gamma^{}\simeq M_2^{}-M_1^{}\ll M_1^{}$\, and
\begin{eqnarray}
{\cal G}(x,y) \,\,=\,\, \int_0^1\frac{du\;u(u - 1)}{u^2x-(1+x-y)u+1} ~,
\end{eqnarray}
in agreement with the expression in the literature~\cite{Schmidt:2012yg}.
Thus Eq.\,(\ref{xrayconstraint}) translates into restrictions on the Yukawa couplings
of $N_{1,2}$.
We remark that with ${\cal Y}_{k1}^{}$ and ${\cal Y}_{k2}^{}$ given by Eq.\,(\ref{123})
the sum in Eq.\,(\ref{N2toN1g}) would vanish if the charged leptons were degenerate.

For the $M_{1,2}$ values considered here,\footnote{Their numbers in our illustrations
in the next section lead to \,$(M_2-M_1)/(M_1+M_2)\lesssim 10^{-8}$.\,
Such a tiny mass split may be explained by the presence of an extra
symmetry, {\it e.g.} particle number conservation, which allows $N_1$ and $N_2$ to form
a pseudo-Dirac fermion, but which is slightly broken by highly suppressed
operators~\cite{Schmidt:2012yg,DeSimone:2010tf}.}
the $N_2$ lifetime \,$\tau_{N_2}^{}=1/\Gamma_{N_2}^{}$\, is dominated by the three body decay
\,$N_2^{}\to N_1^{}\nu\nu$\, which is mediated by the neutral scalars $\cal S$ and $\cal P$
and therefore depends also on ${\cal Y}_{k1,k2}^{}$.
We employ the amplitude and rate already derived in Ref.\,\cite{Ho:2013spa}.

With both $N_1$ and $N_2$ making up the relic density, its observed value constitutes
another restraint on their couplings.
We impose \,$0.1155\le\Omega\hat h^2\le0.1241$\, which is the 90\%\, confidence level (CL)
range of the data \,$\Omega\hat h^2=0.1198\pm0.0026$\,~\cite{pdg,planck},
where $\Omega$ is the present DM density relative to its critical value and $\hat h$ denotes
the Hubble parameter.

Due to the near degeneracy and mutual interactions of $N_{1,2}$, their coannihilation
becomes relevant to the calculation of the relic density~\cite{Griest:1990kh}.
In that case $\Omega$ is approximately given by~\cite{Griest:1990kh,Jungman:1995df}
\begin{eqnarray} \label{omega}
\Omega \hat h^2 &\,=\,& \frac{1.07\times10^9\;x_f^{}{\rm\;GeV}^{-1}}
{\sqrt{g_*^{}}\;m_{\rm Pl}^{}\,\bigl[a_{\rm eff}^{}
+ 3 \bigl(b_{\rm eff}^{}-a_{\rm eff}^{}/4\bigr)/x_f^{}\bigr]} ~,
\nonumber \\
x_f^{} &\,=\,&  \ln\frac{0.191\,\bigl(a_{\rm eff}^{}+6b_{\rm eff}^{}/x_f^{}\bigr)
M_{1\,}^{}m_{\rm Pl}^{}}{\sqrt{g_*^{}\,x_f^{}}} ~,
\end{eqnarray}
where \,$m_{\rm Pl}^{}=1.22\times10^{19}$\,GeV\,~is the Planck mass, $g_*^{}$ is the number of
relativistic degrees of freedom below the freeze-out temperature~\,$T_f^{}=M_1^{}/x_f^{}$, and
$a_{\rm eff}^{}$ and $b_{\rm eff}^{}$ are defined by the expansion of the coannihilation rate
\,$\sigma_{\rm eff}^{}v_{\rm rel}^{}=a_{\rm eff}^{}+b_{\rm eff}^{}v_{\rm rel}^2$\,
in terms of the relative speed $v_{\rm rel}^{}$ of the annihilating particles in their
center-of-mass frame.
The leading contributions to $\sigma_{\rm eff}^{}$ arise from (co)annihilations into
\,$\nu_i^{}\nu_j^{}$\, and \,$\ell_i^-\ell_j^+$,\, which are induced at tree level by
$({\cal S,P})$ and $H^\pm$ exchanges, respectively.
Neglecting the $N_{1,2}$ mass difference, we have~\cite{Griest:1990kh}
\begin{eqnarray}
\sigma_{\rm eff}^{} \,\,=\,\,
\mbox{$\frac{1}{4}$}\bigl(\sigma_{11}^{}+2\sigma_{12}^{}+\sigma_{22}^{}\bigr) ~,
\end{eqnarray}
where \,$\sigma_{kl}^{}=\sigma_{N_k N_l\to\nu\nu}+\sigma_{N_k N_l\to\ell\bar\ell}$\,
and \,$\sigma_{12}^{}=\sigma_{21}^{}$.\,
These cross sections have been computed in Refs.\,\cite{Ho:2013hia,Ho:2013spa} and each
proceed from diagrams in the $t$ and $u$ channels because of the Majorana nature of
the external neutral fermions.
The size of the S-wave contribution $a_{\rm eff}^{}$ is at least several times that of
the P-wave one $b_{\rm eff}^{}$ and comes mainly from $\sigma_{12}^{}$.
In numerical work, we keep both $a_{\rm eff}^{}$ and $b_{\rm eff}^{}$ in~Eq.\,(\ref{omega}).

There are also constraints on ${\cal Y}_{jk}^{}$ from the measurements of a number of
low-energy observables.
These couplings enter the neutrino masses $m_{1,2,3}^{}$ in Eq.\,(\ref{m1m2m3})
and consequently need to be consistent with the most precise mass measurements.
Thus we require
\begin{eqnarray} \label{Delta2}
30.0 \,\,<\,\, \frac{\Delta m^2}{\delta m^2} \,\,<\,\, 34.3
\end{eqnarray}
based on the 90\%\,CL ranges of the data on $\delta m^2$ and $\Delta m^2$ in
Eq.\,(\ref{nudata}).

Another loop process is the flavor-changing radiative decay \,$\mu\to e\gamma$\,
which involves internal $H^\pm$ and $N_k$.
Searches for this decay have come up empty so far, leading to a restraint on
its branching ratio~\cite{pdg},
\begin{eqnarray} \label{mu2eg}
{\cal B}(\mu\to e\gamma)_{\rm exp} \,\,<\,\, 5.7\times10^{-13}
\end{eqnarray}
at 90\% CL.
Hence this caps the prediction~\cite{Kubo:2006yx,Ma:2001mr}
\begin{eqnarray}
{\cal B}(\mu\to e\gamma) \,\,=\,\, \frac{3\alpha}{64\pi\,G_{\rm F}^2\,m_H^4}
\Bigl|\raisebox{0.5ex}{\footnotesize$\displaystyle\sum_k$}\,{\cal Y}_{1k\,}^{}{\cal Y}_{2k}^*\,
{\cal F}\bigl(M_k^2/m_H^2\bigr) \Bigr|^2 ~,
\end{eqnarray}
where $G_{\rm F}$\, is the Fermi constant,
\begin{eqnarray}
\alpha \,\,=\,\, \frac{e^2}{4\pi} ~, \hspace{5ex}
{\cal F}(x) \,\,=\,\, \frac{1-6x+3x^2+2x^3-6x^2\,\ln x}{6(1-x)^4} ~.
\end{eqnarray}

The flavor-diagonal counterpart of the preceding process induces a modification to
the anomalous magnetic moment $a_{\ell_i}$ of lepton $\ell_i$ given by~\cite{Ma:2001mr}
\begin{eqnarray} \label{g-2}
\Delta a_{\ell_i}^{} \,\,=\,\, \frac{-m_{\ell_i}^2}{16\pi^2m_H^2}\,
\raisebox{0.5ex}{\footnotesize$\displaystyle\sum_k$}\,|{\cal Y}_{ik}|^2\,
{\cal F}\bigl(M_k^2/m_H^2\bigr) ~.
\end{eqnarray}
Among $a_{e,\mu,\tau}^{}$, the most restrictive on the potential scotogenic effects
is $a_\mu^{}$ whose current SM and experimental values differ by nearly three sigmas,
\,$a_\mu^{\rm exp}-a_\mu^{\rm SM}=(249\pm87)\times10^{-11}$~\cite{Aoyama:2012wk}.
Accordingly, we require
\begin{eqnarray} \label{gmu-2}
\bigl|\Delta a_\mu^{}\bigr| \,\,<\,\, 9\times10^{-10} ~.
\end{eqnarray}

In addition to the constraints just mentioned, there are others, such as those on
\,$\tau\to(e,\mu)\gamma$,\, as well as theoretical ones, which turn out to be less important
in what follows.
They were described in Ref.\,\cite{Ho:2013hia}.
Direct searches for DM may also add to the restrictions~\cite{Schmidt:2012yg}, but for the
examples below we find that the cross sections of $N_1$ scattering off nuclei can evade
the strongest limits from the LUX experiment~\cite{Akerib:2013tjd}.

After setting $\theta_{12,23,13}$ and $\delta$ to their central values from Eq.\,(\ref{nudata}),
taking \,$E_\gamma=3.54$\,keV\, based on the detected X-ray energy numbers
in~Refs.\,\cite{Bulbul:2014sua,Boyarsky:2014jta,Boyarsky:2014ska},
and scanning the parameter space of the model, we obtain regions satisfying the restrictions
discussed above.
We illustrate this in Table$\;$\ref{bench} with different sets of the mass parameters $m_{0,H}^{}$,
\,$M_1^{}=M_2^{}-E_\gamma^{}$,\, and $M_3^{}$ and the Yukawa constants $Y_{1,2,3}^{}$.
It is worth noting that these results yield
\,$\tau_{N_2}^{}\simeq(\mbox{1.4\,-17})_{\,}\tau_U^{}$\, and
\,$\Gamma_{N_2^{}\to N_1^{}\nu\nu}\simeq(\mbox{12\,-\,84})_{\,}\Gamma_{N_2^{}\to N_1^{}\gamma}$.\,
We turn next to the resulting predictions for a number of observables.

\begin{table}[h]
\caption{Sample values of the mass parameters $m_{0,H}^{}$, \,$M_1\simeq M_2$,\, and $M_3$
and Yukawa constants $Y_{1,2,3}$ satisfying the constraints discussed in
Section~\ref{constraints}.\label{bench}\medskip}
\footnotesize \begin{tabular}{|c|ccccccc|} \hline
\,Set\, & \, $\frac{\displaystyle m_0^{\vphantom{\int}}}{\scriptstyle\rm GeV}$ & \,
$\frac{\displaystyle m_H^{}}{\scriptstyle\rm GeV_{\vphantom{\int}}}$ & ~
$\frac{\displaystyle M_1^{}}{\scriptstyle\rm GeV}$ &
$\frac{\displaystyle M_3^{}}{\scriptstyle\rm GeV}$ ~ & $Y_1^{}$ & $Y_2^{}$ & $Y_3^{}$ \, \\
\hline\hline  I &\,
 340 & \,  395 & ~ 180 & 235 \,& ~ $0.215+0.028i$ \,& ~ $0.281+0.036i$ \,& 0.419 \,
\\  II &\,
 420 & \,  440 & ~ 318 & 415 \,& ~ $0.215+0.027i$ \,& ~ $0.281+0.035i$ \,& 0.431 \,
\\ III &\,
 605 & \,  600 & ~ 350 & 470 \,& ~ $0.120+0.244i$ \,& ~ $0.157+0.319i$ \,& 0.535 \,
\\  IV &\,
1030 & \, 1100 & ~ \, 600 \, & 805 \,&  $-0.360+0.041i$ \,&  $-0.471+0.053i$ \,& 0.716 \,
\\   V &\,
1100 & \, 1200 & ~ 600 & 795 \,&  $-0.377+0.072i$ \,&  $-0.493+0.093i$ \,& 0.750 \,
\\ \hline \end{tabular}
\end{table}

\section{Predictions\label{predictions}}

It is interesting that, although the \,$\mu\to e\gamma$\, bound in Eq.\,(\ref{mu2eg})
is one of the strictest constraints described in the last section, the benchmark points
in Table$\;$\ref{bench} can translate into a branching ratio ${\cal B}(\mu\to e\gamma)$
that is not very close to the experimental limit.
We display the numbers in the second column of Table$\;$\ref{predict}.
Thus they serve as predictions of the scotogenic scenario under consideration that can
be tested with upcoming searches for  \,$\mu\to e\gamma$\,  which will expectedly
reach a sensitivity at a level of a few times $10^{-14}$ within the next five
years~\cite{Cavoto:2014qoa}.

\begin{table}[t]
\caption{Predictions corresponding to the benchmark points in Table~\ref{bench}.
The last three columns contain cross sections at $e^+e^-$ center-of-mass (c.m.) energies
\,$\sqrt s=1,2,3$ TeV.\label{predict}\medskip} \footnotesize
\begin{tabular}{|c|ccccc|ccc|ccc|} \hline
\multirow{2}{*}{\small\,Set\,} &
\multirow{2}{*}{\small$\frac{{\cal B}(\mu\to e\gamma)}{10^{-13}}$} &
\multirow{2}{*}{$\frac{\displaystyle\langle m_{\beta\beta}\rangle}{\scriptstyle\rm eV}$} &
\, \multirow{2}{*}{$\displaystyle\frac{\displaystyle\Sigma_k^{}m_k^{}}{\scriptstyle\rm eV}$} \, &
\, \multirow{2}{*}{$\displaystyle\frac{\alpha_1^{}}{\pi}$} &
\multirow{2}{*}{$\displaystyle\frac{\alpha_2^{}}{\pi}$} &
\multirow{2}{*}{$\frac{\displaystyle\mu_2^{}}{\scriptstyle\rm GeV}$} &
\multirow{2}{*}{\footnotesize${\cal R}_{\gamma\gamma}^{}$} &
\multirow{2}{*}{\footnotesize${\cal R}_{\gamma Z}^{}$} &
\multicolumn{3}{c|}{$\sigma_{e\bar e\to H\bar H\to\ell\bar\ell'_{~}\slash\!\!\!\!E}$
\scriptsize(pb)} \\ \cline{10-12} & & & & & & & & & \scriptsize1 & \scriptsize2 & \scriptsize3 \\
\hline\hline  I$\vphantom{\int^|}$ &
5.6 & 0.054 & 0.20 &     $-0.058$ &         ~ 0.15 &     101 (439)    & 0.91 (1.02) &    0.96 (1.01) &
\, 0.038 \, & 0.059 & \, 0.039 \, \\  II$\vphantom{\int^|}$ &
2.7 & 0.052 & 0.19 &     $-0.061$ &         ~ 0.15 &     110 (499)    & 0.91 (1.03) &    0.96 (1.01) &
0.010 & 0.038 & 0.029 \\ III$\vphantom{\int^|}$ &
3.4 & 0.050 & 0.18 &       ~ 0.57 &         ~ 0.78 &     145 (665)    & 0.91 (1.02) & \, 0.96 (1.01) \, &
    0 & 0.060 & 0.055 \\  IV$\vphantom{\int^|}$ &
0.93 & 0.049 & 0.18 &     $-0.21$ &        ~ 0.001 &     255 (990)    & 0.91 (0.98) &    0.96 (0.99) &
    0 &     0 & 0.054 \\   V$\vphantom{\int_|^|}$ &
0.69 & 0.052 & 0.19 & \,$-0.26$\, & \, $-0.047$ \, & \, 280 (1070) \, & 0.91 (0.98) &    0.96 (0.99) &
    0 &     0 & 0.047 \\
\hline \end{tabular}
\end{table}

Another important observable is the effective Majorana mass
\begin{eqnarray} \label{mee}
\bigl\langle m_{\beta\beta}\bigr\rangle \,\,=\,\, \Bigl|
\raisebox{3pt}{\footnotesize$\displaystyle\sum_k$}\,{\cal U}_{1 k\,}^2 m_k^{} \Bigr|
\,\,=\,\,
\Bigl| c_{12\,}^2 c_{13\,}^2 m_1^{}\,e^{i\alpha_1} + s_{12\,}^2 c_{13\,}^2 m_2^{}\,e^{i\alpha_2}
+ s_{13\,}^2 m_3^{}\,e^{-2i\delta} \Bigr|
\end{eqnarray}
which follows from the Majorana nature of the electron neutrino and
can be probed in neutrinoless double-$\beta$ decay experiments~\cite{nureview}.
This process is of fundamental importance because it violates lepton-number conservation and
thus will be evidence for new physics if detected~\cite{nureview}.
The parameters in Table$\;$\ref{bench} lead to the predictions in the third column of
Table$\;$\ref{predict}.
They are only a few times less than the existing experimental upper limits
on~$\langle m_{\beta\beta}\rangle$, the best one being~\,0.12$\;$eV~\cite{xmbb}.
Forthcoming searches within the next decade are expected to have sensitivities to
$\langle m_{\beta\beta}\rangle$ down to~0.01$\;$eV~\cite{Vignati:2014cqa}.

The sum of neutrino masses, $\Sigma_k^{}m_k^{}$, is also predicted in Table$\;$\ref{predict}.
The results are compatible with the aforementioned bounds from cosmological data,
\,$\Sigma_k^{}m_k^{}<0.18$\,-\,0.23$\;$eV\,~\cite{planck,Riemer-Sorensen:2013jsa},
as well as the hints of greater masses \,$\Sigma_k^{}m_k^{}\sim0.2$\,-\,0.4$\;$eV\,
from other cosmological observations~\cite{Hou:2012xq,Beutler:2014yhv}.
Upcoming data with improved precision can be expected to check the predictions.

We include in Table$\;$\ref{predict} the corresponding values of $\alpha_{1,2}^{}$,
computed using Eq.\,(\ref{alpha12}).
Although currently there is no experimental information on the values of these phases, they may
be extractable from future measurements, especially those on $\langle m_{\beta\beta}\rangle$.

Additional windows into the nonstandard sector of the model may be the Higgs boson
decays \,$h\to\gamma\gamma$\, and \,$h\to\gamma Z$,\, which arise in the SM mainly from
top-quark- and $W$-boson-loop diagrams and also receive one-loop contributions from~$H^\pm$.
Employing the formulas given in Ref.\,\cite{Ho:2013hia} with the Higgs mass
\,$m_h^{}=125.1$\,GeV,\, the average of the most recent
measurements~\cite{Aad:2014aba,CMS:2014ega}, and selecting specific values
of the parameter $\mu_2^{}$ in Eq.\,(\ref{potential}), we have listed in
Table$\;$\ref{predict} the resulting ratio
\begin{eqnarray}
{\cal R}_{\gamma{\cal V}^{\scriptscriptstyle0}}^{} \,\,=\,\,
\frac{\Gamma(h\to\gamma{\cal V}^{\scriptscriptstyle0})}
{\Gamma(h\to\gamma{\cal V}^{\scriptscriptstyle0})_{\rm SM}^{}} ~, \hspace{5ex}
{\cal V}^{\scriptscriptstyle0} \,\,=\,\, \gamma, Z ~,
\end{eqnarray}
where $\Gamma(h\to\gamma{\cal V}^{\scriptscriptstyle0})_{\rm SM}^{}$ is the SM rate.
The two numbers on each line in the ${\cal R}_{\gamma{\cal V}^{\scriptscriptstyle0}}^{}$
column correspond to the two numbers on the same line in the $\mu_2^{}$ column.
Evidently the scotogenic effects on these two modes have a positive correlation.
Since \,$h\to\gamma\gamma$\, has been observed, we can already compare our examples with the data.
The latest measurements of its signal strength performed by the ATLAS and CMS Collaborations
are \,$\sigma/\sigma_{\rm SM}^{}=1.17\pm0.27$\,~\cite{Aad:2014eha} and
\,$\sigma/\sigma_{\rm SM}^{}=1.14^{+0.26}_{-0.23}$\,~\cite{Khachatryan:2014ira}, respectively.
These are both compatible with the predictions, but the situation may
change when more data become available.

As investigated in Ref.\,\cite{Ho:2013spa}, next-generation $e^+e^-$ colliders, such as
the International Linear Collider~\cite{ilc} and Compact Linear Collider~\cite{clic},
have the potential to provide extra means to check the scotogenic model further.
Their c.m. energies may be as high as 3$\;$TeV or more~\cite{ilc}.
Here we consider the scattering \,$e^+e^-\to H^+H^-$\, followed by the (sequential) decays
of $H^\pm$ into \,$\ell_j^\pm N_1^{}$\, possibly plus neutrinos.
Since $N_1$ is DM and the neutrinos are undetected, this process contributes to the channel
\,$e^+e^-\to\ell^+\ell^{\prime-\;}\slash\!\!\!\!E$\, with missing energy \,$\slash\!\!\!\!E$,\,
summed over the final charged leptons.
Using the pertinent expressions derived in Ref.\,\cite{Ho:2013spa}, we collect the scotogenic
contributions in the last three columns of Table$\;$\ref{predict} for c.m. energies
\,$\sqrt s=1,2,3\;$TeV,\, respectively.
The main background is the SM scattering \,$e^+e^-\to W^+W^-\to\nu\nu'\ell^+\ell^{\prime-}$\,
summed over all of the final leptons.
Compared with the SM tree-level cross-sections
\,$\sigma_{e\bar e\to W\bar W\to\nu\nu'\ell\bar\ell'}=0.28,\,0.10,\,0.05\;$pb\,
at \,$\sqrt s=1,2,3$~TeV,\, respectively, clearly the scotogenic numbers can be similar in
size and hence are testable at these colliders.

\section{Conclusions\label{summary}}

We have explored possible implications of X-ray lines from DM-dominated
objects as potential effects of the scotogenic model.
As an example, we concentrate on the unidentified emission line at an energy of
{\small\,$\sim$\,}3.5\,keV\,
recently reported in the X-ray spectra of some galaxy clusters.
Assuming that this finding can stand future scrutiny and that no better standard explanations
are available for it, we consider the scenario in which the line
originates from the decay of fermionic DM in the model.
Specifically, DM is composed of nearly degenerate $N_1$ and~$N_2$, with the latter being
slightly more massive, and the slow decay \,$N_2\to N_1\gamma$\, is responsible for
the observed X-ray line.
Accordingly, we apply various restraints on the model, especially those from the observed DM
relic abundance, neutrino oscillation data, searches for flavor-violating decays such
as~\,$\mu\to e\gamma$,\, and measurements of muon $g$$-$2.
Subsequently, from the allowed parameter space, we pick several benchmark points to make
predictions on a~number of interesting observables in processes involving leptons.
These include the effective Majorana mass that can be probed in ongoing and planned searches
for neutrinoless double-beta decay, the sum of neutrino masses that can be compared with
numbers inferred from upcoming cosmological measurements with improved precision,
the \,$\mu\to e\gamma$\, branching-ratio that will confront further experimental checks not
too long from now, and the rates of the Higgs decays \,$h\to\gamma\gamma$\,
and~\,$h\to\gamma Z$\, presently being investigated at the~LHC.
Many of the predictions are already within reach of current or near future experiments.
We also evaluate the cross section of
\,$e^+e^-\to H^+H^-\to\ell^+\ell^{\prime-\;}\slash\!\!\!\!E$\, which is testable
at next-generation $e^+e^-$ colliders.
These machines can, in addition, provide a~cleaner environment than the LHC to
measure\,$\;h\to\gamma\gamma,\gamma Z$.\,
Thus our analysis indicates that X-ray lines from certain astronomical objects can potentially
offer extra means to examine the scotogenic model.

\acknowledgments

This research was supported in part by the MOE Academic Excellence Program (Grant No. 102R891505),
the research grant NTU-ERP-102R7701, and the NCTS.


\newpage

\begin{thebibliography}{0}

\bibitem{Bulbul:2014sua}
  E.~Bulbul, M.~Markevitch, A.~Foster, R.K.~Smith, M.~Loewenstein, and S.W.~Randall,
  Astrophys.\ J.\  {\bf 789}, 13 (2014)
  [arXiv:1402.2301 [astro-ph.CO]].

\bibitem{Boyarsky:2014jta}
  A.~Boyarsky, O.~Ruchayskiy, D.~Iakubovskyi, and J.~Franse,
  arXiv:1402.4119 [astro-ph.CO].

\bibitem{Boyarsky:2014ska}
  A.~Boyarsky, J.~Franse, D.~Iakubovskyi and O.~Ruchayskiy,
  arXiv:1408.2503 [astro-ph.CO].

\bibitem{Riemer-Sorensen:2014yda}
  S.~Riemer-S{\o}rensen,
  arXiv:1405.7943 [astro-ph.CO];
  T.E.~Jeltema and S.~Profumo,
  arXiv:1408.1699 [astro-ph.HE];
  D.~Malyshev, A.~Neronov, and D.~Eckert,
  arXiv:1408.3531 [astro-ph.HE];
M.E.~Anderson, E.~Churazov, and J.N.~Bregman,
  arXiv:1408.4115 [astro-ph.HE].

\bibitem{Boyarsky:2014paa}
  A.~Boyarsky, J.~Franse, D.~Iakubovskyi, and O.~Ruchayskiy,
  arXiv:1408.4388 [astro-ph.CO].

\bibitem{f2fg}
  H.~Ishida, K.S.~Jeong, and F.~Takahashi,
  Phys.\ Lett.\ B {\bf 732}, 196 (2014)  [arXiv:1402.5837 [hep-ph]];
  D.P.~Finkbeiner and N.~Weiner,
  arXiv:1402.6671 [hep-ph];
  T.~Higaki, K.S.~Jeong, and F.~Takahashi,
  Phys.\ Lett.\ B {\bf 733}, 25 (2014)  [arXiv:1402.6965 [hep-ph]];
  J.~Jaeckel, J.~Redondo, and A.~Ringwald,
  Phys.\ Rev.\ D {\bf 89}, 103511 (2014)  [arXiv:1402.7335 [hep-ph]];
  K.N.~Abazajian,
  Phys.\ Rev.\ Lett.\  {\bf 112}, 161303 (2014)  [arXiv:1403.0954 [astro-ph.CO]];
  R.~Krall, M.~Reece, and T.~Roxlo,
  arXiv:1403.1240 [hep-ph];
  C.~El Aisati, T.~Hambye, and T.~Scarna,
  JHEP {\bf 1408}, 133 (2014)  [arXiv:1403.1280 [hep-ph]];
  S.~Baek and H.~Okada,
  arXiv:1403.1710 [hep-ph];
  K.Y.~Choi and O.~Seto,
  Phys.\ Lett.\ B {\bf 735}, 92 (2014)  [arXiv:1403.1782 [hep-ph]];
  M.~Cicoli, J.P.~Conlon, M.C.D.~Marsh, and M.~Rummel,
  Phys.\ Rev.\ D {\bf 90}, 023540 (2014)  [arXiv:1403.2370 [hep-ph]];
  F.~Bezrukov and D.~Gorbunov,
  arXiv:1403.4638 [hep-ph];
  C.~Kolda and J.~Unwin,
  Phys.\ Rev.\ D {\bf 90}, 023535 (2014)  [arXiv:1403.5580 [hep-ph]];
  R.~Allahverdi, B.~Dutta, and Y.~Gao,
  Phys.\ Rev.\ D {\bf 89}, 127305 (2014)  [arXiv:1403.5717 [hep-ph]];
  A.G.~Dias, A.C.B.~Machado, C.C.~Nishi, A.~Ringwald, and P.~Vaudrevange,
  JHEP {\bf 1406}, 037 (2014)  [arXiv:1403.5760 [hep-ph]];
  N.E.~Bomark and L.~Roszkowski,
  Phys.\ Rev.\ D {\bf 90}, 011701 (2014)  [arXiv:1403.6503 [hep-ph]];
  S.P.~Liew,
  JCAP {\bf 1405}, 044 (2014)  [arXiv:1403.6621 [hep-ph]];
  Z.~Kang, P.~Ko, T.~Li, and Y.~Liu,
  arXiv:1403.7742 [hep-ph];
  F.S.~Queiroz and K.~Sinha,
  Phys.\ Lett.\ B {\bf 735}, 69 (2014)  [arXiv:1404.1400 [hep-ph]];
  H.~Okada and T.~Toma,
  Phys.\ Lett.\ B {\bf 737}, 162 (2014)  [arXiv:1404.4795 [hep-ph]];
  H.M.~Lee,
  arXiv:1404.5446 [hep-ph];
  D.J.~Robinson and Y.~Tsai,
  Phys.\ Rev.\ D {\bf 90}, 045030 (2014)  [arXiv:1404.7118 [hep-ph]];
  J.P.~Conlon and F.V.~Day,
  arXiv:1404.7741 [hep-ph];
  K.~Nakayama, F.~Takahashi, and T.T.~Yanagida,
  Phys.\ Lett.\ B {\bf 737}, 311 (2014)  [arXiv:1405.4670 [hep-ph]];
  S.~Chakraborty, D.K.~Ghosh, and S.~Roy,
  arXiv:1405.6967 [hep-ph];
  M.~Lattanzi, R.A.~Lineros, and M.~Taoso,
  arXiv:1406.0004 [hep-ph];
  J.P.~Conlon and A.J.~Powell,
  arXiv:1406.5518 [hep-ph];
  C.Q.~Geng, D.~Huang, and L.H.~Tsai,
  JHEP {\bf 1408}, 086 (2014)  [arXiv:1406.6481 [hep-ph]];
  A.~Abada, G.~Arcadi, and M.~Lucente,
  arXiv:1406.6556 [hep-ph];
  H.~Baer, K.Y.~Choi, J.E.~Kim, and L.~Roszkowski,
  arXiv:1407.0017 [hep-ph];
  H.~Okada and Y.~Orikasa,
  arXiv:1407.2543 [hep-ph];
  W.~Rodejohann and H.~Zhang,
  Phys.\ Lett.\ B {\bf 737}, 81 (2014)  [arXiv:1407.2739 [hep-ph]];
  N.~Haba, H.~Ishida, and R.~Takahashi,
  arXiv:1407.6827 [hep-ph];
  C.W.~Chiang and T.~Yamada,
  JHEP {\bf 1409}, 006 (2014)  [arXiv:1407.0460 [hep-ph]];
  J.M.~Cline and A.R.~Frey,
  arXiv:1408.0233 [hep-ph];
  B.~Henning, J.~Kehayias, H.~Murayama, D.~Pinner, and T.T.~Yanagida,
  arXiv:1408.0286 [hep-ph];
  Y.~Farzan and A.R.~Akbarieh,
  arXiv:1408.2950 [hep-ph];
  T.~Higaki, N.~Kitajima, and F.~Takahashi,
  arXiv:1408.3936 [hep-ph].

\bibitem{Modak:2014vva}
  K.P.~Modak,
  arXiv:1404.3676 [hep-ph].

\bibitem{Frandsen:2014lfa}
  M.T.~Frandsen, F.~Sannino, I.M.~Shoemaker, and O.~Svendsen,
  JCAP {\bf 1405}, 033 (2014)  [arXiv:1403.1570 [hep-ph]].

\bibitem{ann}
  E.~Dudas, L.~Heurtier, and Y.~Mambrini,
  Phys.\ Rev.\ D {\bf 90}, 035002 (2014)
  [arXiv:1404.1927 [hep-ph]];
  S.~Baek, P.~Ko, and W.I.~Park,
  arXiv:1405.3730 [hep-ph].

\bibitem{Ma:2006km}
  E.~Ma,
  Phys.\ Rev.\ D {\bf 73}, 077301 (2006)  [hep-ph/0601225].

\bibitem{Kubo:2006yx}
  J.~Kubo, E.~Ma, and D.~Suematsu,
  Phys.\ Lett.\ B {\bf 642}, 18 (2006)
  [hep-ph/0604114].

\bibitem{Sierra:2008wj}
D.~Aristizabal Sierra, J.~Kubo, D.~Restrepo, D.~Suematsu, and O.~Zapata,
  Phys.\ Rev.\ D {\bf 79}, 013011 (2009)  [arXiv:0808.3340 [hep-ph]];
  D.~Suematsu, T.~Toma, and T.~Yoshida,
  Phys.\ Rev.\ D {\bf 79}, 093004 (2009)  [arXiv:0903.0287 [hep-ph]];
  G.B.~Gelmini, E.~Osoba, and S.~Palomares-Ruiz,
  Phys.\ Rev.\ D {\bf 81}, 063529 (2010)  [arXiv:0912.2478 [hep-ph]];
  R.~Bouchand and A.~Merle,
  JHEP {\bf 1207}, 084 (2012)  [arXiv:1205.0008 [hep-ph]];
  E.~Ma,
  Phys.\ Lett.\ B {\bf 717}, 235 (2012)  [arXiv:1206.1812 [hep-ph]];
  P.K.~Hu,
  arXiv:1208.2613 [hep-ph];
  S.~Bhattacharya, E.~Ma, A.~Natale, and A.~Rashed,
  Phys.\ Rev.\ D {\bf 87}, 097301 (2013)  [arXiv:1302.6266 [hep-ph]];
  T.~Toma and A.~Vicente,
  JHEP {\bf 1401}, 160 (2014)  [arXiv:1312.2840 [hep-ph]].

\bibitem{Schmidt:2012yg}
  D.~Schmidt, T.~Schwetz, and T.~Toma,
  Phys.\ Rev.\ D {\bf 85}, 073009 (2012)  [arXiv:1201.0906 [hep-ph]].

\bibitem{Ho:2013hia}
  S.Y.~Ho and J.~Tandean,
  Phys.\ Rev.\ D {\bf 87}, 095015 (2013)  [arXiv:1303.5700 [hep-ph]].

\bibitem{Ho:2013spa}
  S.Y.~Ho and J.~Tandean,
  Phys.\ Rev.\ D {\bf 89}, 114025 (2014)  [arXiv:1312.0931 [hep-ph]].

\bibitem{pdg}
  K.~A.~Olive {\it et al.}  [Particle Data Group Collaboration],
  Chin.\ Phys.\ C {\bf 38}, 090001 (2014).

\bibitem{pmns}
  B.~Pontecorvo,
Sov.\ Phys.\ JETP {\bf 26} (1968) 984
  [Zh.\ Eksp.\ Teor.\ Fiz.\  {\bf 53} (1968) 1717];
  Z.~Maki, M.~Nakagawa, and S.~Sakata,
  Prog.\ Theor.\ Phys.\  {\bf 28}, 870 (1962).

\bibitem{Capozzi:2013csa}
  F.~Capozzi, G.L.~Fogli, E.~Lisi, A.~Marrone, D.~Montanino, and A.~Palazzo,
  Phys.\ Rev.\ D {\bf 89}, 093018 (2014)  [arXiv:1312.2878 [hep-ph]].

\bibitem{planck}
  P.A.R.~Ade {\it et al.}  [Planck Collaboration],
  Astron.\ Astrophys.\  (2014)  [arXiv:1303.5076 [astro-ph.CO]].

\bibitem{Riemer-Sorensen:2013jsa}
  S.~Riemer-S{\o}rensen, D.~Parkinson, and T.M.~Davis,
  arXiv:1306.4153 [astro-ph.CO].

\bibitem{Hou:2012xq}
Z.~Hou  {\it et al.}, 
  Astrophys.\ J.\  {\bf 782}, 74 (2014)  [arXiv:1212.6267 [astro-ph.CO]].

\bibitem{Beutler:2014yhv}
  F.~Beutler {\it et al.}  [BOSS Collaboration],
  arXiv:1403.4599 [astro-ph.CO].

\bibitem{DeSimone:2010tf}
  A.~De Simone, V.~Sanz, and H.P.~Sato,
  Phys.\ Rev.\ Lett.\  {\bf 105}, 121802 (2010)  [arXiv:1004.1567 [hep-ph]].

\bibitem{Griest:1990kh}
  K.~Griest and D.~Seckel,
  Phys.\ Rev.\  D {\bf 43}, 3191 (1991).

\bibitem{Jungman:1995df}
  G.~Jungman, M.~Kamionkowski, and K.~Griest,
  Phys.\ Rept.\  {\bf 267}, 195 (1996)  [hep-ph/9506380];
  J.~Edsjo and P.~Gondolo,
  Phys.\ Rev.\ D {\bf 56}, 1879 (1997)  [hep-ph/9704361].

\bibitem{Ma:2001mr}
  E.~Ma and M.~Raidal,
  Phys.\ Rev.\ Lett.\  {\bf 87}, 011802 (2001)
  [Erratum-ibid.\  {\bf 87}, 159901 (2001)]  [hep-ph/0102255].

\bibitem{Aoyama:2012wk}
  T.~Aoyama, M.~Hayakawa, T.~Kinoshita, and M.~Nio,
  Phys.\ Rev.\ Lett.\  {\bf 109}, 111808 (2012)
  [arXiv:1205.5370 [hep-ph]].

\bibitem{Akerib:2013tjd}
  D.S.~Akerib {\it et al.}  [LUX Collaboration],
  Phys.\ Rev.\ Lett.\  {\bf 112}, 091303 (2014)
  [arXiv:1310.8214 [astro-ph.CO]].

\bibitem{Cavoto:2014qoa}
  G.~Cavoto,
  arXiv:1407.8327 [hep-ex].

\bibitem{nureview}
For recent reviews, see
  W.~Rodejohann,
  J.\ Phys.\ G {\bf 39}, 124008 (2012)  [arXiv:1206.2560 [hep-ph]];
S.T.~Petcov,
  Int.\ J.\ Mod.\ Phys.\ A {\bf 29}, 1430028 (2014)  [arXiv:1405.2263 [hep-ph]].

\bibitem{xmbb}
A.~Gando {\it et al.}  [KamLAND-Zen Collaboration],
  Phys.\ Rev.\ Lett.\  {\bf 110}, no. 6, 062502 (2013)  [arXiv:1211.3863 [hep-ex]];
  M.~Agostini {\it et al.}  [GERDA Collaboration],
  Phys.\ Rev.\ Lett.\  {\bf 111}, no. 12, 122503 (2013)  [arXiv:1307.4720 [nucl-ex]];
  R.~Arnold {\it et al.}  [NEMO-3 Collaboration],
  Phys.\ Rev.\ D {\bf 89}, 111101 (2014)  [arXiv:1311.5695 [hep-ex]];
  J.B.~Albert {\it et al.}  [EXO-200 Collaboration],
  Nature {\bf 510}, 229–234 (2014)  [arXiv:1402.6956 [nucl-ex]].

\bibitem{Vignati:2014cqa}
  M.~Vignati,
  EPJ Web Conf.\  {\bf 70}, 00044 (2014).

\bibitem{Aad:2014aba}
  G.~Aad {\it et al.}  [ ATLAS Collaboration],
  arXiv:1406.3827 [hep-ex].

\bibitem{CMS:2014ega}
CMS Collaboration,
Report No. CMS-PAS-HIG-14-009, July 2014.

\bibitem{Aad:2014eha}
  G.~Aad {\it et al.}  [ATLAS Collaboration],
  arXiv:1408.7084 [hep-ex].

\bibitem{Khachatryan:2014ira}
  V.~Khachatryan {\it et al.}  [CMS Collaboration],
  arXiv:1407.0558 [hep-ex].

\bibitem{ilc}
  T.~Behnke
{\it et al.},
  arXiv:1306.6327 [physics.acc-ph].

\bibitem{clic}
http://clic-study.org.

\end{thebibliography}
\end{document}